\documentclass[%
 aip,
 jmp,%
amsmath,amssymb,
 reprint,%
]{revtex4-1}

\usepackage{amssymb}
\usepackage[permil]{overpic}
\usepackage{graphicx}
\usepackage{dcolumn}

\begin{document}

 \preprint{AIP/123-QED}

\title{Photon-induced thermal effects in superconducting coplanar waveguide resonators}

\author{Yiwen Wang}
 \affiliation{
Quantum Optoelectronics Laboratory, School of Physics, Southwest
Jiaotong University, Chengdu 610031, China}

\author{Pinjia Zhou}
 \affiliation{
Quantum Optoelectronics Laboratory, School of Physics, Southwest
Jiaotong University, Chengdu 610031, China}

\author{Lianfu Wei}
 \email{weilianfu@gmail.com}
 \affiliation{
Quantum Optoelectronics Laboratory, School of Physics, Southwest
Jiaotong University, Chengdu 610031, China}
 \affiliation{
State Key Laboratory of Optoelectronic Materials and Technologies,
School of Physics and Engineering, Sun Yat-Sen University, Guangzhou
510275, China}

\author{Haijie Li}
 \affiliation{
Quantum Optoelectronics Laboratory, School of Physics, Southwest
Jiaotong University, Chengdu 610031, China}

\author{Beihong Zhang}
 \affiliation{
Quantum Optoelectronics Laboratory, School of Physics, Southwest
Jiaotong University, Chengdu 610031, China}

\author{Miao Zhang}
 \affiliation{
Quantum Optoelectronics Laboratory, School of Physics, Southwest
Jiaotong University, Chengdu 610031, China}

\author{Qiang Wei}
 \affiliation{
Quantum Optoelectronics Laboratory, School of Physics, Southwest
Jiaotong University, Chengdu 610031, China}

\author{Yurong Fang}
 \affiliation{
Research Institute of Superconductor Electronics, Department of
Electronic Science and Engineering, Nanjing University, Nanjing
210093, China}

\author{Chunhai Cao}
 \affiliation{
Research Institute of Superconductor Electronics, Department of
Electronic Science and Engineering, Nanjing University, Nanjing
210093, China}

\begin{abstract}
We experimentally investigated the optical responses of a
superconducting niobium resonator. It was found that, with
increasing radiation power, the resonance frequency
increases monotonically below around 500 mK, decreases
monotonically above around 1 K and exhibits a nonmonotonic behavior at around $700$ mK. These observations
show that one can operate the irradiated resonator in three temperature regimes, depending on whether two-level system (TLS) effects or kinetic inductance effects dominate. Furthermore,
we found that the optical responses at ultra-low temperatures
can be qualitatively regarded as a photon-induced thermalization effect of TLSs, which could be utilized to achieve thermal sensitive
photon detections.
\end{abstract}


\maketitle In recent years superconducting coplanar waveguide (CPW)
resonators have attracted extensive attentions, due to its important
applications in solid-state quantum information processings and
sensitive photon detections~\cite{pkday1,gao1,Baselmans1,Mazin1}.
Particularly, high-quality factor
resonators~\cite{Barends1,Vissers1,Megrant1} (typically $Q >
10^{5}$) operating well below transition temperature $T_{c}$ have
been demonstrated to be suitable for serving as so-called microwave
kinetic inductance detectors (MKIDs)~\cite{Zmuidzinas}. According to
Mattis-Bardeen theory~\cite{Zmuidzinas2}, the absorption of photons
(or increasing bath temperature) will decrease the electron-pair
density in superconducting film, causing an increase in the kinetic
inductance~\cite{Mazin2} and thus a decrease in the resonance
frequency. Meanwhile, the dissipation increases due to quasiparticle
excitations, leading to a reduction in the quality factor. Therefore
by probing the changes in the frequency or dissipation of a MKID,
the radiation light signals can be detected. Much work to date have
been done to optimize resonator geometries and materials as well as
a better understanding of the noise properties in order to improve
the detection sensitivity of MKID~\cite{gao2,Barends2,Leduc}.

In this paper we try to demonstrate another possible photon
detection approach with a superconducting resonator. This approach
is based on the thermalization of two-level systems (TLSs) in
dielectrics rather than the kinetic inductance effects in the metal
film. With a high-quality factor niobium resonator, we studied its
optical responses by performing transmission measurements at low
temperatures in a broad range of $20$ mK to $1.9$ K, well below the
niobium transition temperature $T_{c} = 9.2$ K. We found that the
irradiated niobium resonator can be operated in three temperature
regimes: with increasing radiation power, the resonance frequency
increases monotonically below $\sim 500$ mK, decreases monotonically
above $\sim 1$ K and exhibits an interesting nonmonotonicity at
temperatures around $700$ mK.  Similar crossover behavior with bath
temperature for a unirradiated resonator have been
reported~\cite{gao3,Barends3,gao4,Lindstrom,gao5}. Our observations
can be qualitatively explained by the temperature dependent
permittivity (due to TLSs in dielectrics) and temperature dependent
kinetic inductance. At sufficiently low temperatures, TLS effects
dominate over kinetic inductance effects, suggesting that the
photon-induced thermalization effects of TLSs in dielectrics could
be utilized to achieve thermal sensitive photon detections. Note
that this is a different photon detection approach from MKID, which
typically requires a superconducting material and geometry with high
kinetic inductance fraction~\cite{Porch}.

Our CPW resonator~\cite{Wang} was fabricated by magnetron sputtering
and photolithography. A $160$ nm thick niobium film was deposited on
a $500$ $\mu$m thick high-resistivity silicon substrate. Contact
exposure as well as reflect ion etching were used to generate
desired patterns on the film. The microscope photograph of the key
structure of the measured sample is shown in Fig.~1(a). The
resonator is actually a meandered $\lambda /4$ transmission line,
which shorts to the ground plane at one end and capacitively couples
to a feedline at another end. The feedline is $20$ $\mu$m wide with
a gap of $15$ $\mu$m to the ground plane. The center line of the
resonator is $7$ $\mu$m wide with a gap of $7$ $\mu$m. The total
length of the resonator is about $15.814$ mm and the relative
permittivity of Si substrate is $11.9$, giving a theoretical
fundamental resonance frequency $f_{r} = 1.8665$ GHz.

\begin{figure}[tbp]
\begin{center}
\begin{overpic}[width=8.2cm]{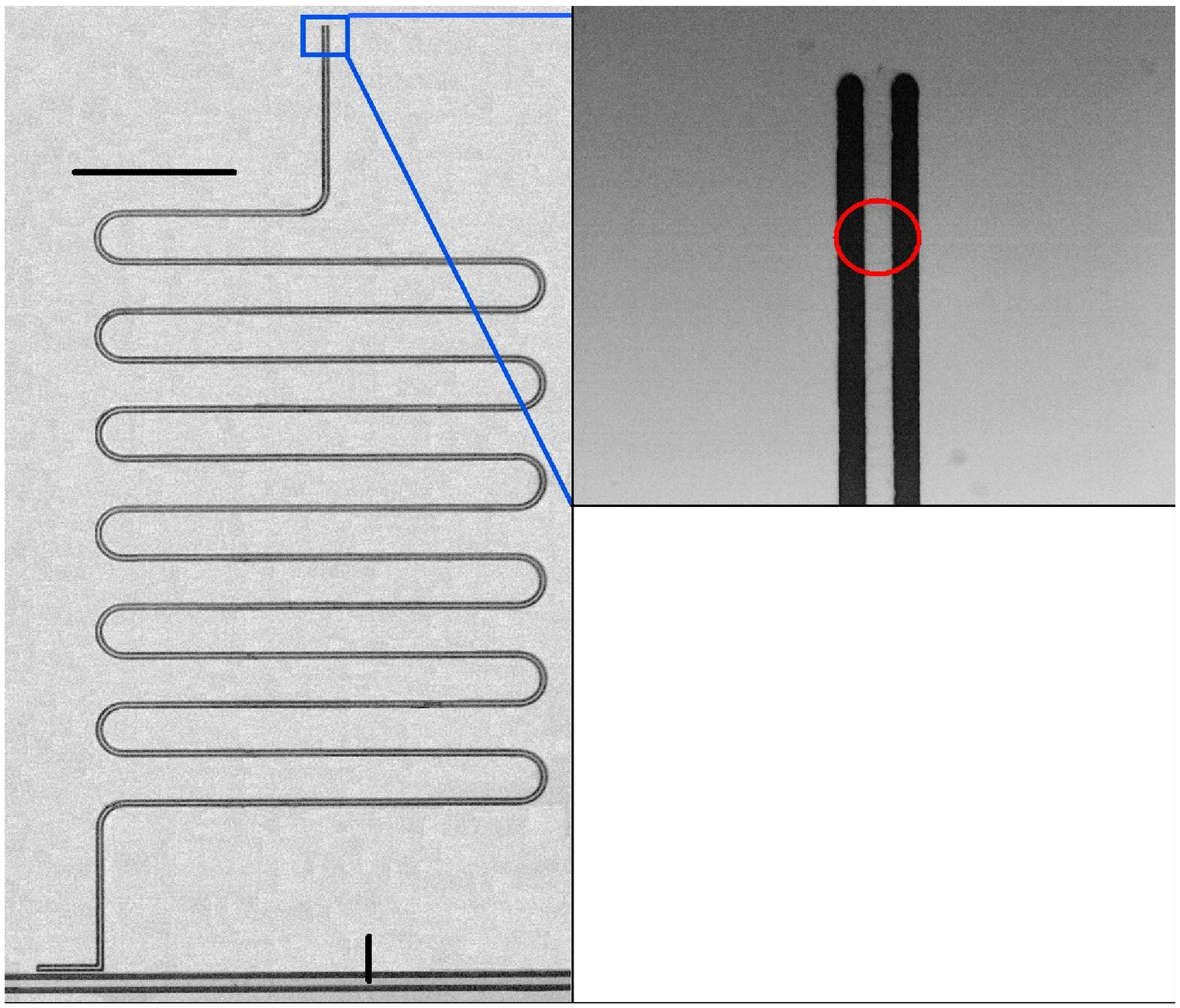}
\put(18,800){\bf (a)} \put(500,800){\bf (b)} \put(485,390){\bf (c)}
\put(68,730){$\footnotesize 400\mu m$} \put(210,70){Feedline}
\put(123,389){CPW Resonator} \put(906,800){Nb}
\put(490,5){\includegraphics[width=4.35cm]{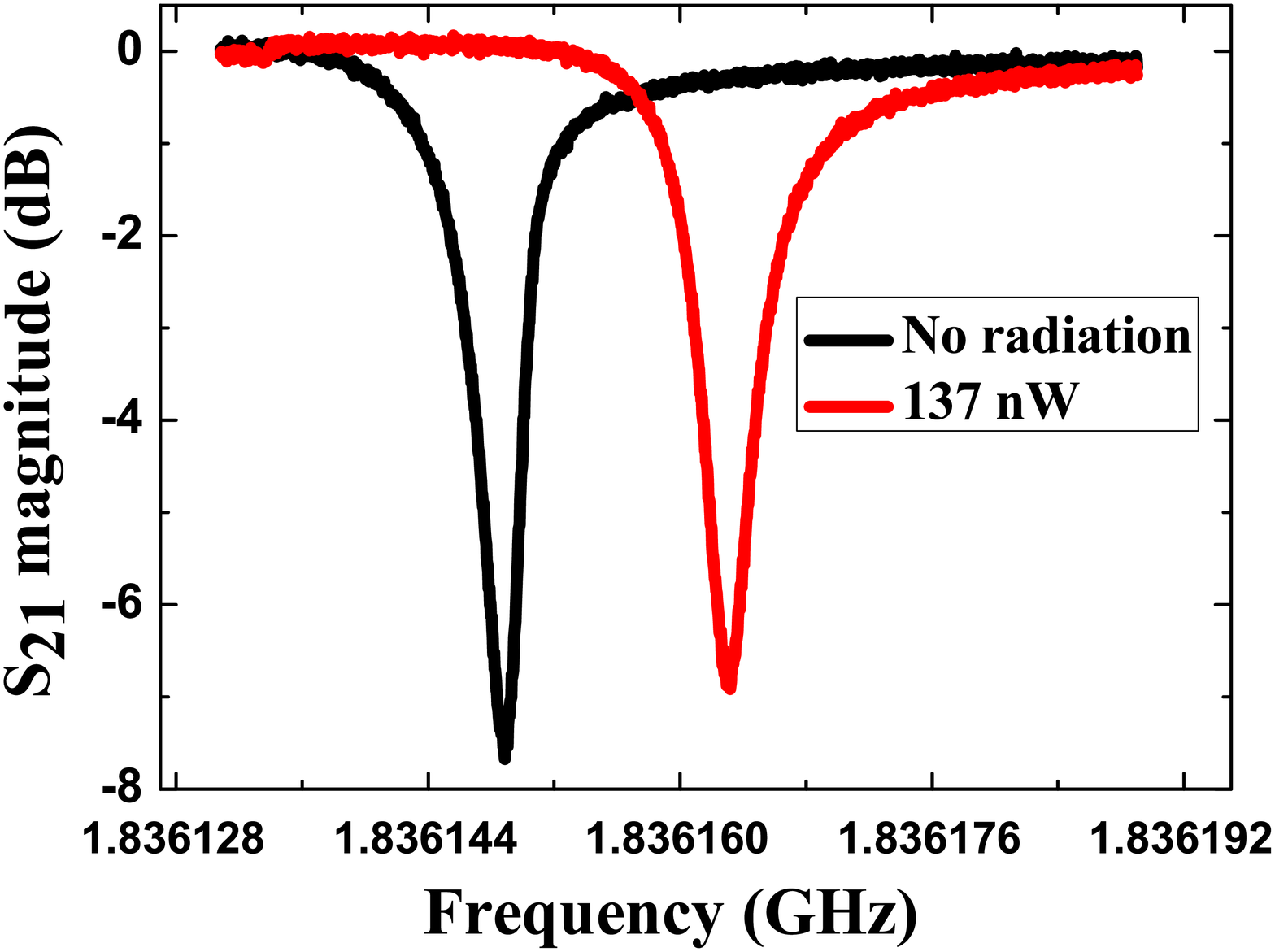}}
\end{overpic}
\end{center}
\small {\caption{(Color online) (a) Microscope photograph of the
measured resonator. Light and dark areas are niobium film and
silicon substrate respectively. The meandered resonator is
capacitively coupled to a feedline, which is used for microwave
excitation and readouts. The center line of the resonator is $7$
$\mu$m wide with a gap of $7$ $\mu$m to ground plane. (b) The short
end area of resonator. The red circle is about $20$ $\mu$m in
diameter, indicating the area where the photons are incident on. (c)
The observed resonance dips. The black curve shows the transmission
data with no radiation, corresponding to a resonance frequency
$f_{r1} = 1.836149$ GHz and loaded quality factor $Q_{1}$ = $3.7$
$\times 10^{5}$. In the presence of a steady radiation with power
$137$ nW, the resonance frequency increases to $f_{r2} = 1.836163$
GHz and the quality factor decreases to $Q_{2}$ = $2.8$ $\times
10^{5}$, as seen by the red curve.}}
\end{figure}

The resonator chip is glued and wire-bonded to a gold-plated
oxygen-free copper box equipped with microwave connections. This
sample box is mounted at the cold plate of the mixing chamber in a
dilution refrigerator with base temperature less than $20$ mK. The
temperature was measured using a calibrated R1101 resistance
thermometer which is placed close to the sample box. To radiate the
CPW resonator, a single-mode optical fiber is set up from the room
temperature down to the sample box. A room temperature laser source
is connected to the top end of the fiber and generates a steady
stream of photons at wavelength $635$ nm. The bottom end of the
fiber is carefully aligned and fixed so that the incoming photons
can hit the short end area of the resonator, where the standing wave
current distribution is maximum. The fiber end is about $100$ $\mu$m
vertically away from the chip surface and the irradiation area
(indicated by the red circle in Fig.~1(b)) is about $20$ $\mu$m in
diameter. Therefore, part of the incoming photons are incident on
the superconducting film and part on the bare substrate in the gap.
Additionally, not all the photons are absorbed by the chip, as some
photons could be reflected and scattered into the environment.

The device transmission amplitudes $S_{21}$ as a function of
frequency were measured with an Agilent E5071C vector network
analyzer. Attenuators and DC blocks are positioned appropriately to
suppress circuit noises. The microwave driving power reaching the
chip is estimated to be about $-65$ dBm, which is fixed for all the
measurements in this letter. At temperature $20$ mK, a resonance dip
was found at $f_{r1} = 1.836149$ GHz, shown in Fig.~1(c). Note that
the resonance curves have been calibrated to measure $0$ dB off
resonance to remove the effects of attenuation in the circuits. By
fitting the $\lvert S_{21}\lvert ^{2}$ data to a skewed Lorentzian
model~\cite{Gao-phd}, one can obtain the loaded quality factor of
the resonator $Q_{1}$ = $3.7$ $\times 10^{5}$. We then radiated the
resonator with a power $137$ nW and measured the transmissions when
the system reaches thermal and electric equilibrium. As shown in
Fig.~1(c), the resonance frequency shifts to a higher value $f_{r2}
= 1.836163$ GHz and the quality factor decreases to $Q_{2}$ = $2.8$
$\times 10^{5}$. The decrease in quality factor is expected but the
increase in resonance frequency is in contrary to the prediction of
Mattis-Bardeen theory.


We now investigate the optical responses of the resonator with
different radiation powers. To this aim we varied the light
intensity and measured corresponding microwave transmissions through
the resonator at base temperature $T$ = $20$ mK. Fig.~2 (blue
circles) shows the relative resonance frequency shift as a function
of the radiation power on the short end area. Here, the relative
frequency shift is defined as $\Delta f_{r}/f_{r} = [f_{r}(P) -
f_{r}(P=0)]/f_{r}(P=0)$, with $P$ being the total radiation power
entering the sample box and $f_{r}(P)$ the corresponding resonance
frequency. It is shown that the frequency shift increases
monotonically with increasing radiation power in the range of $0.4$
nW to $2.2$ $\mu$W. The inset shows the same plot except the
power-axis is in logarithmic scale. One can see that the frequency
shift increases approximately linearly with logarithmic power above
a crossover point (at $\approx $ $1$ nW), indicated by
the intersection of two red dotted lines.

\begin{figure}[tbp]
\includegraphics[width=7.5cm]{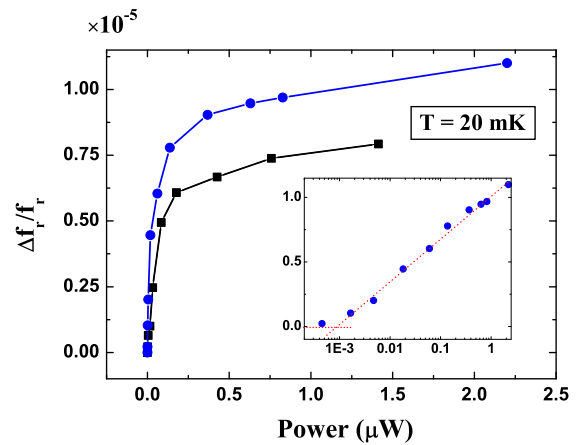}
\small { \caption{The relative resonance frequency shift as a
function of radiation power at $T =$ $20$ mK. The blue circles
(connected by lines as guides to eyes) correspond to radiation on
short end area. The inset plots the same data with logarithmic
power. The horizontal red dotted line corresponds to $\Delta
f_{r}/f_{r} = 0$ and the oblique one shows the linear dependence on
the logarithmic power. The black squares correspond to radiation on
bare substrate near the resonator, exhibiting a similar power
dependence but with a lower responsivity.}}
\end{figure}

Qualitatively, the radiation power dependence observed here is
similar to the temperature dependence reported in a few recent
experiments~\cite{gao3,Barends3,gao4,Lindstrom,gao5}, where the
measured resonance frequency shift shows approximately a logarithmic
increase with temperature well below $T_{c}$. Such behaviors were
also verified in our measurements of the relative frequency shift
versus bath temperature, as shown in Fig.~3.
\begin{figure}[tbp]
\begin{center}
\begin{overpic}[width=7.0cm]{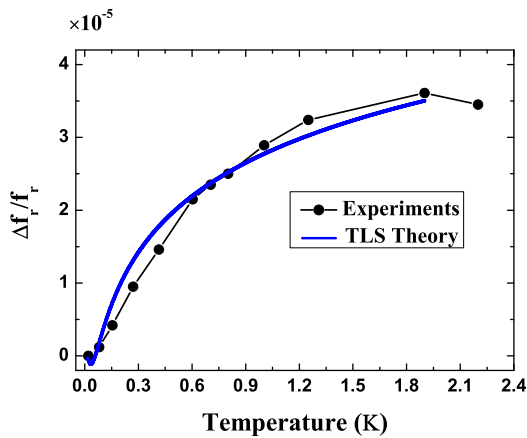}
\end{overpic}
\end{center}
\small {\caption{(Color online) Relative resonance frequency shift
$\Delta f_{r}/f_{r}$ as a function of temperature in the absence of
radiation. The black solid lines are guides to eyes. Fits to the TLS
theory (blue line) agree well with the experimental data below $1.3$
K.}}
\end{figure}
Here, $\Delta f_{r}/f_{r} = (f_{r}(T) -f_{r}(T_{0}))/f_{r}(T_{0})$
with the reference temperature $T_{0} =$ $20$ mK. Note that the
experiments can not be explained by Mattis-Bardeen theory, which
predicts a frequency shift opposite in sign. Instead, the observed
temperature dependence can be attributed to the TLS effects in
dielectrics~\cite{Martinis2,Martinis3}. In fact, TLSs are abundant
in amorphous materials and can be present in substrate, the surface
oxides on the metal film and the interface between metal and
substrate~\cite{Martinis}. At low temperatures, the unsaturated TLSs
with electric dipole moments can interact resonantly with the
microwave field, giving a temperature dependent variation of
permittivity~\cite{Phillips}. Changing in permittivity affects the
capacitance per unit length and thus the resonance frequency.
Assuming a uniform distribution of TLSs, the relative resonance
frequency shift is given by~\cite{gao3}
\begin{equation}
\frac{\Delta
f_{r}}{f_{r}}=C\left\{\ln\left(\frac{T}{T_{0}}\right)-\left[g(T,\omega)-g(T_{0},\omega)\right]\right\},
\end{equation}
where $C$ is the unique fitting parameter, $T_{0} =$ $20$ mK the
reference temperature and $g(T,\omega)
=\text{Re}\left[\Psi(\frac{1}{2}+\hbar\omega/2\pi ik_BT)\right]$
with $\Psi$ being the complex digamma function. Fits to the above
equation ($C =$ $1.15$ $\times 10^{-5}$) can explain the observed
temperature dependence below $1.9$ K, where the resonance frequency
increases monotonically with temperature. This suggests that at low
temperatures $T << T_{c}$, which is the case here, the thermal
effects on the kinetic inductance are relatively
weak~\cite{Meservey}. Additionally, considering the geometry of our
resonator, its kinetic inductance fraction is estimated to be very
small (less than $0.02$). Therefore the kinetic inductance effects
are negligible for $T << T_{c}$ and TLS effects become prominent,
which is the reason why Mattis-Bardeen theory does not describe the
dominating loss mechanism at ultra-low temperatures. However, the
resonance frequency starts to drop at around $\sim 2$ K, indicating
that TLS effects saturate and kinetic inductance becomes dominant as
temperature approaching $T_{c}$. Similar experimental results have
been observed in a NbTiN resonator covered with SiO$_x$
layer~\cite{Barends3}.


We believe that the above TLS effects can also explain the observed
radiation power dependence at $T =$ $20$ mK. Experimentally, photons
irradiated on both metal film and substrate will mainly heat the
chip and increase its effective temperature, although the bath
temperature does not change. At ultra-low temperatures(e.g., $T =$
$20$ mK), the incident photons can excite TLSs in the substrate,
especially those in the exposed substrate surface (i.e., SiO$_x$ in
our device) and nearby the irradiated area. To further verify it is
indeed the thermalization of TLSs in the substrate dominates the
radiation power dependence of resonance frequency at ultra-low
temperatures, we also directly radiated on a small area of the bare
substrate, which is about $5$ mm away from the resonator. We
performed the relevant transmission measurements at $20$ mK and
obtained a similar radiation power dependence, as shown in Fig.~2
(black squares). The similar power dependence implies that the
thermalization of TLSs is the main factor in shifting the resonace
frequency. Moreover, the photon responsivity is weaker when
radiating on the bare substrate of a certain distance away from the
resonator than that when radiating on the short end area. This is
reasonable and can be explained by a nonuniform temperature
distribution around the irradiated area. The effective temperature
at the RF active component of the resonator is lower when the $20$
$\mu$m thermal source is moved $5$ mm away.


\begin{figure}[tbp]
\begin{center}
\begin{overpic}[width=7.5cm]{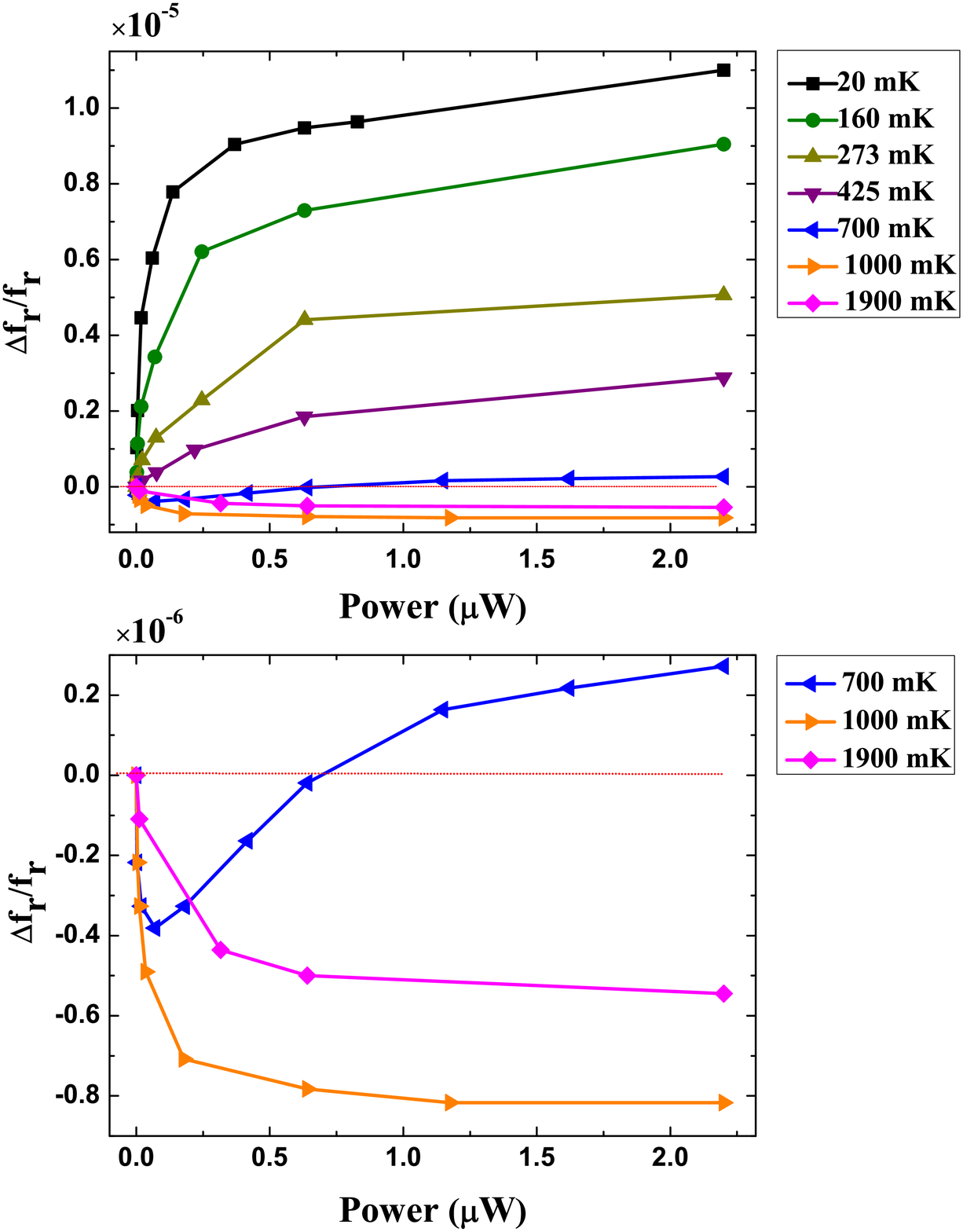}
\put(30,980){\bf (a)} \put(30,480){\bf (b)}
\end{overpic}
\end{center}
\small {\caption{(Color online) (a) Relative resonance frequency
shift $\Delta f_{r}/f_{r}$ as a function of photon power at various
temperatures. The solid lines are guides to eyes. The horizontal red
dotted line indicates $\Delta f_{r}/f_{r} = 0$. (b) The magnified
view of the photon power dependence of resonance frequency at higher
temperatures.}}
\end{figure}

We then investigated the optical responses of the resonator at
different bath temperatures. Fig.~4(a) shows the relative resonance
frequency shift as a function of radiation power, taken at seven
selective bath temperatures in a broad range of $20$ mK - $1900$ mK.
Here, the relative frequency shift is defined as $\Delta f_{r}/f_{r}
= (f_{r}(P,T) - f_{r}(P=0,T))/f_{r}(P=0,T)$, with $T$ being the bath
temperature and $P$ the radiation power. Note that the reference
frequency $f_{r}(P=0,T)$ varies with temperature $T$. At lower
temperatures (i.e., $20$ mK, $160$ mK, $273$ mK and $425$ mK), the
frequency shift increases monotonically with increasing radiation
power, indicating the TLS effects dominate in this temperature
range. Apparently, the resonator responsivity to photons is
strongest at the lowest temperature $T =$ $20$ mK. For the
temperature at $700$ mK, we observed an interesting phenomenon that
the resonance frequency exhibits a nonmonotonic behavior with
increasing radiation power: the frequency firstly goes down and then
goes up, which can be seen more clearly in Fig.~4(b). This suggests
that the kinetic inductance effects start to play in role at the
temperature around $700$ mK and the TLS effects are less pronounced
than those at ultra-low temperatures. With lower radiation power,
the incident photons break Cooper pairs near the irradiated area,
increasing the kinetic inductance and decreasing the resonance
frequency. However, with enhancing radiation power, the chip is
heated to excite the unsaturated TLSs and the TLS effects may become
dominated over the kinetic inductance effects again, causing the
resonance frequency increase. Furthermore, for higher temperatures,
e.g. at $1000$ mK and $1900$ mK, the frequency decreases
monotonically with radiation power. As temperatures increase, the
TLS effects tend to fully saturate. Therefore only kinetic
inductance changes with incident photons in high temperature regime.

In summary, we experimentally studied the optical responses of a
superconducting niobium coplanar waveguide resonator at various
temperatures. Our results show that one can operate an irradiated Nb
resonator in three different temperature regimes below $T_{c}$. At
ultra-low temperatures (below $\sim 500$ mK), TLS effects dominate
while at higher temperatures (above $\sim$ $1$ K), TLS effects
saturate and kinetic inductance effects dominate. In the middle
regime (e.g., around $700$ mK), both TLSs and kinetic inductance
take effects. Therefore, instead of maximizing the kinetic
inductance effects to improve the performance of MKID for sensitive
photon detections, the TLS effects could also be utilized to achieve
weak light detections at ultra-low temperatures. Furthermore, lower
bath temperatures lead to stronger photon responses.

Although the sensitivity of the TLS-based photon detections
demonstrated here is much lower than that of a MKID, the relevant
resonator is relatively easy to be fabricated since high kinetic
inductance fraction is not required. The performance of proposed
TLS-based weak light detections could be improved by minimizing the
heat capacity of the chip. The chip size can be designed and
fabricated as small as possible so that a certain radiation power
can lead to a bigger effective temperature change of the resonator.
Also, an antireflection-coating can be used to reduce photon
reflections and achieve a high optical coupling efficiency. In
addition, increasing the density of TLSs in dielectrics will enhance
the TLS-resonator coupling and thus in principle may improve the
device sensitivity to incident photons.

\vspace{0.4cm}

This work was supported in part by the National Natural Science
Foundation (Grant No. 61301031, No. 11174373, No. 11204249), and the
National Fundamental Research Program of China (Grant No.
2010CB923104). We thank Profs. Y. Yu and P. H. Wu for kind supports
and discussions.

\nocite{*}
\bibliography{aipsamp}

\end{document}